\documentclass[
twocolumn,
amsmath,amssymb,
aps,
]{revtex4-1}
\usepackage{graphicx}% Include figure files
\usepackage{hyperref}
\usepackage{dcolumn}% Align table columns on decimal point
\usepackage{bm}% bold math
\usepackage{epstopdf}
\usepackage{ifpdf}
\usepackage{romannum}

\begin{document}

\preprint{Sun/depairing Jc}

\title{Achieving the depairing limit along $c$ axis in Fe$_{1+y}$Te$_{1-x}$Se$_x$ single crystals}% Force line breaks with \\

\author{Yue Sun$^1$}
\email{sunyue@phys.aoyama.ac.jp}
\author{Haruka Ohnuma$^1$, Shin-ya Ayukawa$^2$, Takashi Noji$^3$, Yoji Koike$^3$, Tsuyoshi Tamegai$^4$}
\author{Haruhisa Kitano$^1$}
\email{hkitano@phys.aoyama.ac.jp}

\affiliation{%
$^1$Department of Physics and Mathematics, Aoyama Gakuin University, Sagamihara 252-5258, Japan\\
$^2$Research Institute for Interdisciplinary Science, Okayama University, Okayama 700-8530, Japan\\
$^3$Department of Applied Physics, Tohoku University, Sendai 980-8579, Japan\\
$^4$Department of Applied Physics, The University of Tokyo,Tokyo 113-8656, Japan}

\date{\today}% It is always \today, today,

\begin{abstract}
We report the achieving of depairing current limit along $c$-axis in Fe$_{1+y}$Te$_{1-x}$Se$_x$ single crystals. A series of crystals with $T_{\rm{c}}$ ranging from 8.6 K to 13.7 K (different amount of excess Fe, $y$) were fabricated into $c$-axis bridges with a square-micrometer cross-section. The critical current density, $J_{\rm{c}}$, was directly estimated from the transport current-voltage measurements. The transport $J_{\rm{c}}$ reaches a very large value, which is about one order of magnitude larger than the depinning $J_{\rm{c}}$, but comparable to the calculated depairing $J_{\rm{c}}$ $\sim$ 2 $\times$ 10$^6$ A/cm$^2$ at 0 K, based on the Ginzburg-Landau (GL) theory. The temperature dependence of the depairing $J_{\rm{c}}$ follows the GL-theory ($\propto$ (1-$T/T_{\rm{c}}$)$^{3/2}$) down to $\sim$ 0.83 $T_{\rm{c}}$, then increases with a reduced slope at low temperatures, which can be qualitatively described by the Kupriyanov-Lukichev theory. Our study provides a new route to understand the behavior of depairing $J_{\rm{c}}$ in iron-based superconductors in a wide temperature range.

%\begin{description}
%\verb+74.70.Xa+, \verb+74.62.En+, \verb+74.25.fc+
%\end{description}
\end{abstract}

\maketitle
\section{introduction}
Iron chalcogenide superconductors have attracted much attention because of the discovery of high temperature superconductivity. Although the superconducting (Sc) transition temperature, $T_c$, in FeSe is only 9 K \cite{HsuFongChiFeSediscovery}, it can be easily enhanced to 14 K by Te substitution \cite{SalesPRB}, up to 37 K under pressure \cite{MedvedevNatMat}, and over 40 K by intercalating spacer layers \cite{BurrardNatMat}. More interestingly, the monolayer of FeSe grown on SrTiO$_3$ shows a large $T_c$ $\sim$ 65 K \cite{HeShaolongNatMat}. Fe$_{1+y}$Te$_{1-x}$Se$_{x}$ is unique in their structural simplicity, consisting of only FeTe/Se layers, which is favorable for probing the superconducting mechanism. Recently, a topological surface superconductivity \cite{ZhangARPESScience,ZhangPengNatPhy}, and the possible Majorana bound state have been observed \cite{WangMajoranaScience,MachidaNatMat}, which make Fe$_{1+y}$Te$_{1-x}$Se$_{x}$ the first high temperature topological superconductor. On the other hand, its high upper critical field ($\sim$ 50 Tesla) and less toxic nature compared with iron pnictides suggest that Fe$_{1+y}$Te$_{1-x}$Se$_x$ are also favorable to applications. Until now, the superconducting tapes with the transport critical current density, $J_{\rm{c}}$, over 10$^6$ A/cm$^2$ under self-field and over 10$^5$ A/cm$^2$ under 30 T at 4.2 K have already been fabricated \cite{SiWeidongNatComm}. 

The transport $J_{\rm{c}}$ determined by the depairing process of Cooper pairs, called depairing $J_{\rm{c}}$, is crucial for the study of the superconducting mechanism, because it directly provides information on the critical velocity of superfluids, and the magnitude as well as the symmetry of the SC gap \cite{Tinkham}. The depairing process occurs when the kinetic energy of the supercurrent exceeds the condensation energy ($\propto$ SC gap) \cite{Tinkham, Dew-HughesJcReview}. However, it is difficult to be achieved, since the vortex flow occurs preceding the depairing at much smaller current density. The critical current density determined by the vortex flow, which occurs when the Lorentz force exceeds the pinning force of vortex, is usually called depinning $J_{\rm{c}}$ \cite{Dew-HughesJcReview,Blatterreview}. The depinning $J_{\rm{c}}$ obtained from the magnetic hysteresis loops (MHLs) by using the Bean model \cite{Beanmodel} is mainly determined by the defects, disorders, and the geometry of the samples \cite{Blatterreview}. On the other hand, $J_{\rm{c}}$ can be also obtained from the direct transport measurements on thin films \cite{DEVRIESJcCupratesfilmReview,H_nisch_SUSTreviewIBSfilm}. The small thickness of thin films allows us to reach the critical limit by applying not so large current. However, single-crystalline thin films without weak links are hard to fabricate due to the presence of grain boundaries and twin boundaries from the growth technique \cite{HilgenkampRevModPhys.74.485,Durrell_reviewGrainBoundary}.

To achieve the depairing $J_{\rm{c}}$, direct transport current-voltage ($I$-$V$) measurement on a single crystal is required. However, it is very difficult for bulk samples to achieve this limit since the extremely large current is needed. To solve this problem,  micro-fabrication technique is used to reduce the size of the crystal to micrometer or sub-micrometer scale. \cite{FioryAPL,SkocpolPhysRevB.14.1045,Xunanplett,RomijnPhysRevB.26.3648,YangPeidongScience,RusanovPhysRevB.70.024510,CirilloPhysRevB.75.174510,MollSmFeAsOFNatMat,NawazPhysRevLettYBCOJdp,LIANGSLCOJdPhysicaC,LiJunAPLdeparingJcBaK122}. Until now, studies on the depairing $J_{\rm{c}}$ have been mainly performed on low-$T_{\rm{c}}$ superconductors, especially at low temperatures  \cite{FioryAPL,SkocpolPhysRevB.14.1045,Xunanplett,RomijnPhysRevB.26.3648,RusanovPhysRevB.70.024510,CirilloPhysRevB.75.174510,Kunchur_JdMgB2Review}. For iron-based superconductors (IBSs), the depairing $J_{\rm{c}}$ has been probed on the micro-fabricated Ba$_{1-x}$K$_x$Fe$_2$As$_2$ single crystal with current flowing in the $ab$-plane \cite{LiJunAPLdeparingJcBaK122}. Nonetheless, it was only performed at the temperatures close to $T_{\rm{c}}$ due to the extremely large in-plane depairing $J_{\rm{c}}$. The depairing $J_{\rm{c}}$ for Ba$_{1-x}$K$_x$Fe$_2$As$_2$ is found to follow the prediction of the GL-theory at temperatures close to $T_{\rm{c}}$ \cite{LiJunAPLdeparingJcBaK122}. However, the behavior of the depairing $J_{\rm{c}}$ at low temperatures for IBSs still remains unknown.     

\begin{table*}
	\caption{Depairing current density for typical iron-based superconductors, calculated by $J_{\rm{dp}}^{\rm{GL}}=c\phi_0/12\sqrt3\pi^2\xi(0)\lambda(0)^2$, where $c$ is the speed of light, $\phi_0$ is the flux quantum, $\xi$(0) is the coherence length at 0 K, $\lambda$(0) is the penetration depth at 0 K, respectively. Here, we should note that the anisotropies of some IBSs estimated from penetration depth $\lambda_c$/$\lambda_{ab}$ and coherence length $\xi_{ab}$/$\xi_c$ are different at temperatures much below $T_{\rm{c}}$ \cite{PROZOROVPhyscaC}. Therefore, we prefer to use the experimental values of $\lambda$ and $\xi$ for the calculations.}
	\begin{ruledtabular}
		\begin{tabular}{c c c c c c c}
			& $\lambda_{ab}$(0) ($\mu$m) & $\lambda_{c}$(0) ($\mu$m) & $\xi_{ab}$(0) (nm) & $\xi_{c}$(0) (nm) & $J_{\rm{dp}}^{\rm{GL}}$($ab$) (A/cm$^2$) & $J_{\rm{dp}}^{\rm{GL}}$($c$) (A/cm$^2$)   \\	
			FeTe$_{1-x}$Se$_x$ & 0.49 \cite{BendelePhysRevB_lambda}  &1.32 \cite{BendelePhysRevB_lambda}  & 2.8 \cite{LeiHechangPhysRevcoherencelenth} & 3 \cite{LeiHechangPhysRevcoherencelenth} & 1.5 $\times$ 10$^7$  & 2.0 $\times$ 10$^6$ \\
			FeSe & 0.45 \cite{Kasahara18112014} & 1.1 \cite{CaoAIPAdvance} & 4.3 \cite{LeiHechangPhysRevB.84.014520}& 2.9 \cite{VedeneevPhysRevB.87.134512}   & 1.2 $\times$ 10$^7$  & 2.9 $\times$ 10$^6$ \\
			NdFeAsO$_{1-x}$F$_x$ & 0.2 \cite{PROZOROVPhyscaC}  & 3.7 \cite{PROZOROVPhyscaC} & 2.3 \cite{Jia_NdFeAsO_APL}& 0.26 \cite{Jia_NdFeAsO_APL} & 1.1 $\times$ 10$^8$   & 2.8 $\times$ 10$^6$\\
			Ba$_{1-x}$K$_x$Fe$_2$As$_2$ &0.25 \cite{PROZOROVPhyscaC}  &1.8 \cite{PROZOROVPhyscaC}  & 2.2 \cite{Yuannature} & 2.2 \cite{Yuannature}  & 7.4 $\times$ 10$^7$   & 1.4 $\times$ 10$^6$\\
			Ba(Fe$_{1-x}$Co$_x$)$_2$As$_2$ & 0.2 \cite{Prozorov_Review_Penetration}  & 1.0 \cite{Prozorov_Review_Penetration} & 2.4 \cite{YamamotoBaCo122APL} & 1.2 \cite{YamamotoBaCo122APL} & 1.1 $\times$ 10$^8$   & 8.4 $\times$ 10$^6$ \\
			BaFe$_2$(As$_{1-x}$P$_x$)$_2$ & 0.2 \cite{HashimotoScience}  & -  & 3.2 \cite{CHONG20101178SSC} & 1.3 \cite{CHONG20101178SSC}  & 7.9 $\times$ 10$^7$  & -\\
			K$_x$Fe$_2$Se$_2$ & 0.29 \cite{TorchettiKFe2Se2Penetration} &  - & 2.3 \cite{MunPhysRevB.83.100514} & 1.4$\sim$2.3 \cite{MunPhysRevB.83.100514}  &  5.2 $\times$ 10$^7$ & -\\
			LiFeAs & 0.21 \cite{InosovPhysRevLett.104.187001} & -  & 4.8 \cite{ZhangLiFeAsPhysRevB.83.174506} &1.7 \cite{ZhangLiFeAsPhysRevB.83.174506}  & 4.8 $\times$ 10$^7$  & -

		\end{tabular}
	\end{ruledtabular}
\end{table*}

To reveal the behavior of the depairing $J_{\rm{c}}$ at lower temperatures, we turn eyes on the $J_{\rm{c}}$ along $c$-axis. As summarized in Table \Romannum{1}, the theoretical depairing $J_{\rm{c}}$ along $c$-axis for IBSs is about one order smaller than that in the $ab$-plane due to the larger penetration depth $\lambda_{\rm{c}}$, which is advantageous for achieving the depairing limit. Until now, the $J_{\rm{c}}$ along $c$-axis in micro-fabricated single crystals has already been studied in IBSs (V$_2$Sr$_4$O$_6$)Fe$_2$As$_2$ and Sm/PrFeAs(O,F). However, the depairing limit is not achieved because of the possible formation of intrinsic Josephson junctions \cite{MollNatPhy,MollSmFeAsOFNatMat,Kashiwayaapl}. In this study, we focus on the Fe$_{1+y}$Te$_{1-x}$Se$_{x}$ single crystals, whose depinning $J_{\rm{c}}$ for both $H\parallel c$ and $H\parallel ab$ are typically 3 $\times$ 10$^5$ A/cm$^2$ at 2 K \cite{SunAPEX,Sun_SUSTevolution}. When the current is flowing in the $ab$-plane, the depairing $J_{\rm{c}}$ is estimated as 1.5 $\times$ 10$^7$ A/cm$^2$ (see Table \Romannum{1}). On the other hand, when the current is flowing along $c$-axis, the depairing $J_{\rm{c}}$ is estimated as 2.0 $\times$ 10$^6$ A/cm$^2$ (see Table \Romannum{1}), which is more suitable for probing the depairing $J_{\rm{c}}$ at low temperatures. 

In this report, we successfully achieve the depairing limit in Fe$_{1+y}$Te$_{1-x}$Se$_{x}$ single crystals by fabricating narrow bridges along $c$-axis. The obtained transport $J_{\rm{c}}$ is about one order of magnitude larger than the depinning $J_{\rm{c}}$, indicating the obtained $J_{\rm{c}}$ is the depairing $J_{\rm{c}}$. Besides, the temperature dependence of the depairing $J_{\rm{c}}$ is studied down to $\sim$ 0.3 $T_{\rm{c}}$, which can be qualitatively described by the Kupriyanov-Lukichev (KL) theory.

\section{experiment}
Fe$_{1+y}$Te$_{1-x}$Se$_x$ ($x$ = 0.2, 0.3, and 0.4) single crystals were grown by slow cooling method \cite{NojiFeTeSeannealing,SunSUSTFeSeannealing}. All the crystals show plate-like morphology, and can grow up to the size of centimeter. Only the (00$l$) peaks are observed in the single crystal x-ray diffraction, suggesting that the crystallographic $c$-axis is perfectly perpendicular to the plane of the single crystal \cite{SunSciRep2016}. Crystal composition is evaluated by the inductively-coupled plasma (ICP) atomic emission spectroscopy, which confirms that the actual Se-doping level is very close to the nominal one \cite{NojiFeTeSeannealing,SunSciRep2016}. On the other hand, the as-grown crystals usually contain some amount of excess Fe (represented by $y$, $\sim$ 0.14 in the Fe$_{1+y}$Te$_{0.6}$Se$_{0.4}$ single crystal \cite{SunSciRep}) residing in the interstitial sites of the Te/Se layer. The excess Fe was removed and its amount was tuned by both the post annealing and electrochemical reaction method, as reported in our previous publications \cite{SunSciRep,OkadaJJAP}. Together with the removing of excess Fe, $T_{\rm{c}}$ is found to be spontaneously increased \cite{SunSciRep,SunPRB2014}. In the fully-annealed crystals, the excess Fe was totally removed as been confirmed by the scanning tunneling microscopy (STM) measurements \cite{MachidaNatMat,SunSciRep}. By these methods, a series of single crystals with different $T_{\rm{c}}$, i.e. different amount of excess Fe, were prepared. More details about the crystal, excess Fe, and the basic properties can be seen in a recent review paper \cite{Sun_sustreview}. 

\begin{figure}\center
	\includegraphics[width=8.5cm]{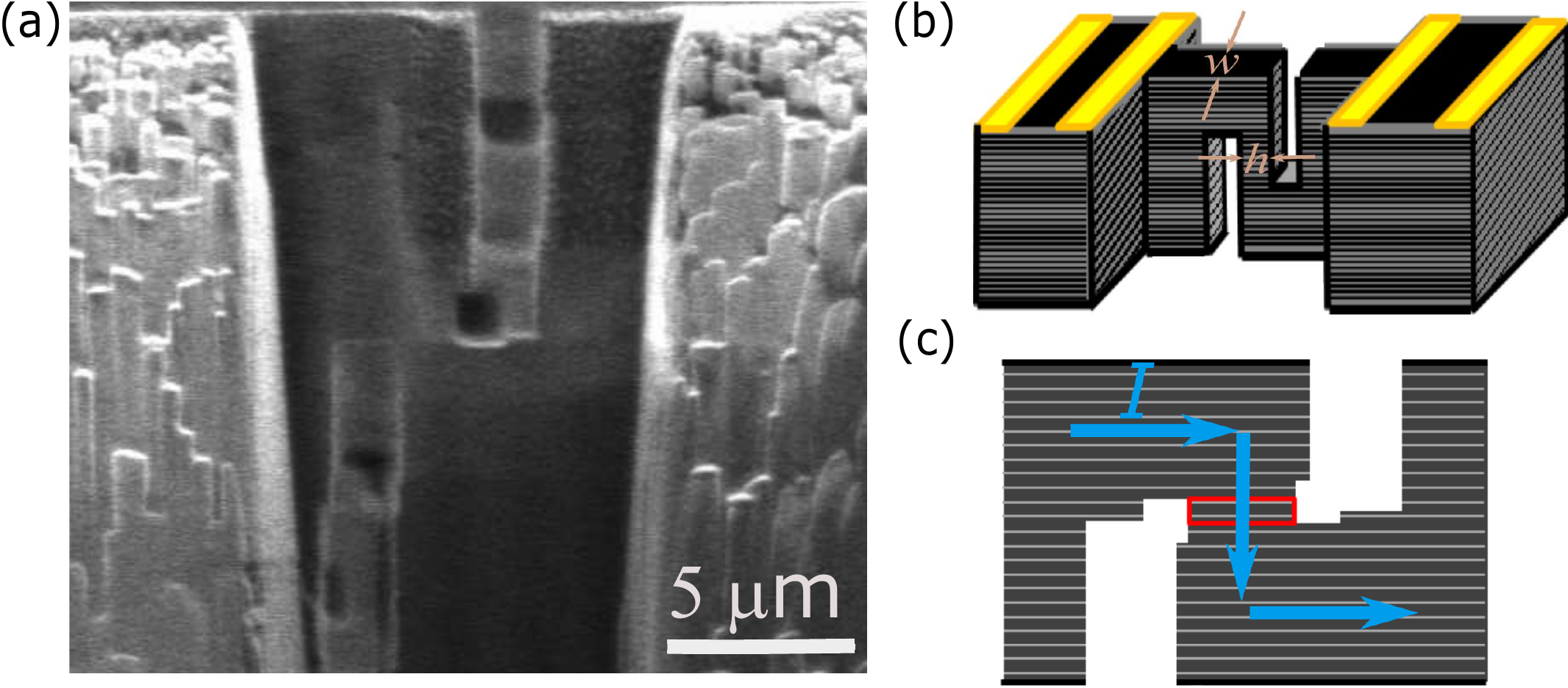}
	\caption{(a) The scanning ion microscopy images of the fabricated $c$-axis structure in Fe$_{1+y}$Te$_{1-x}$Se$_x$ single crystal. Schematic drawing of (b) the bridge structure, and the (c) current flowing path along the $c$-axis bridge.}\label{}
\end{figure}

\begin{table}
	\caption{Cross-section areas ($w$ $\times$ $h$) of the fabricated $c$-axis bridges. $\varLambda$ is the Pearl length calculated by 2$\lambda_c^2/h$ \cite{ClemPearllength}, where $\lambda_c$ is the penetration depth along $c$-axis, $\sim$ 1.6 $\mu$m at 5 K \cite{BendelePhysRevB_lambda}.}
	\begin{ruledtabular}
		\begin{tabular}{c c c c c }
			$T_{\rm{c}}$ (K)  & $w$ ($\mu$m) & $h$ ($\mu$m) & $\Lambda$ ($\mu$m)   \\	
			8.6    & 2.0  & 1.1      & 4.7   \\
			8.8    & 2.0  & 1.1      & 4.7   \\
			9.3    & 1.4  & 1.3      & 3.9  \\
			9.5    & 1.4  & 1.0      & 5.1   \\
			10.9   & 0.8  & 1.2      & 4.3  \\
			11.9   & 1.7  & 1.5      & 3.4 \\
			12.8   & 1.4  & 0.9      & 5.7 \\
			13.7   & 3.4  & 1.0      & 5.1

		\end{tabular}
	\end{ruledtabular}
\end{table}

The $c$-axis bridge, as shown schematically in Fig. 1(b), was fabricated by using the focused ion beam (FIB) technique \cite{KakehiIEEE,Kakizaki_JJAP,Ayukawa}. The single crystal was first cleaved into a slice with a thickness smaller than 10 $\mu$m, and fixed on a sapphire substrate. Then the central part of the crystal was necked down to a length of 5 $\sim$ 10 $\mu$m and a width of $\sim$ 1 $\mu$m from the top by FIB. The necked part was further fabricated into two separated slits with a typical overlap of 100 $\sim$ 200 nm, which will enforce the current to flow along $c$-axis in the bridge region as marked by the rectangular frame in Fig. 1(c). With such a structure, the critical current density of the device is determined by the $c$-axis bridge since the other parts of the crystal have much larger cross area. The scanning ion microscopy image of a typical $c$-axis bridge structure is shown in Fig. 1(a). The cross-section areas ($w$ $\times$ $h$) of the bridges together with the value of $T_{\rm{c}}$ are listed in Table \Romannum{2}.

Resistance measurements were performed by a standard four-probe method. The $I$-$V$ measurements were performed below $T_{\rm{c}}$ by applying two kinds of pulse currents to the $c$-axis bridge. In the first method, the bias current was linearly swept up and down through a standard resistance (1 k$\Omega$) by using an arbitrary-waveform generator (Agilent 33220A). The width of such a ramped current pulse is 5 ms, and the repetition period is 143 ms. In the second method, the pulse current was applied by using a Keithley Delta Pulse System. Rectangular 100 $\mu$s current pulses were passed through the sample at 3 s intervals (duty ratio $\sim$ 3.3 $\times$ 10$^{-5}$) to avoid heating effect. The voltage drop across the bridge was integrated for 55 $\mu$s. To avoid damage to the $c$-axis bridge, the pulse current was stopped when the voltage reaches the threshold value of 30 $\mu$V.

Here, we point out that the resistance measurements probe the region intervening between the two voltage contacts on the crystal, where the narrow bridge and other crystalline parts are included. On the other hand, the $I$-$V$ measurements performed below $T_{\rm{c}}$ probe only the part of micro-fabricated narrow bridge. Thus, the effects of sample inhomogeneity in the resistance measurements should be carefully distinguished with those in the $I$-$V$ measurements, as will be discussed in the next section.

\section{results and discussion}

\begin{figure}\center
	\includegraphics[width=8.5cm]{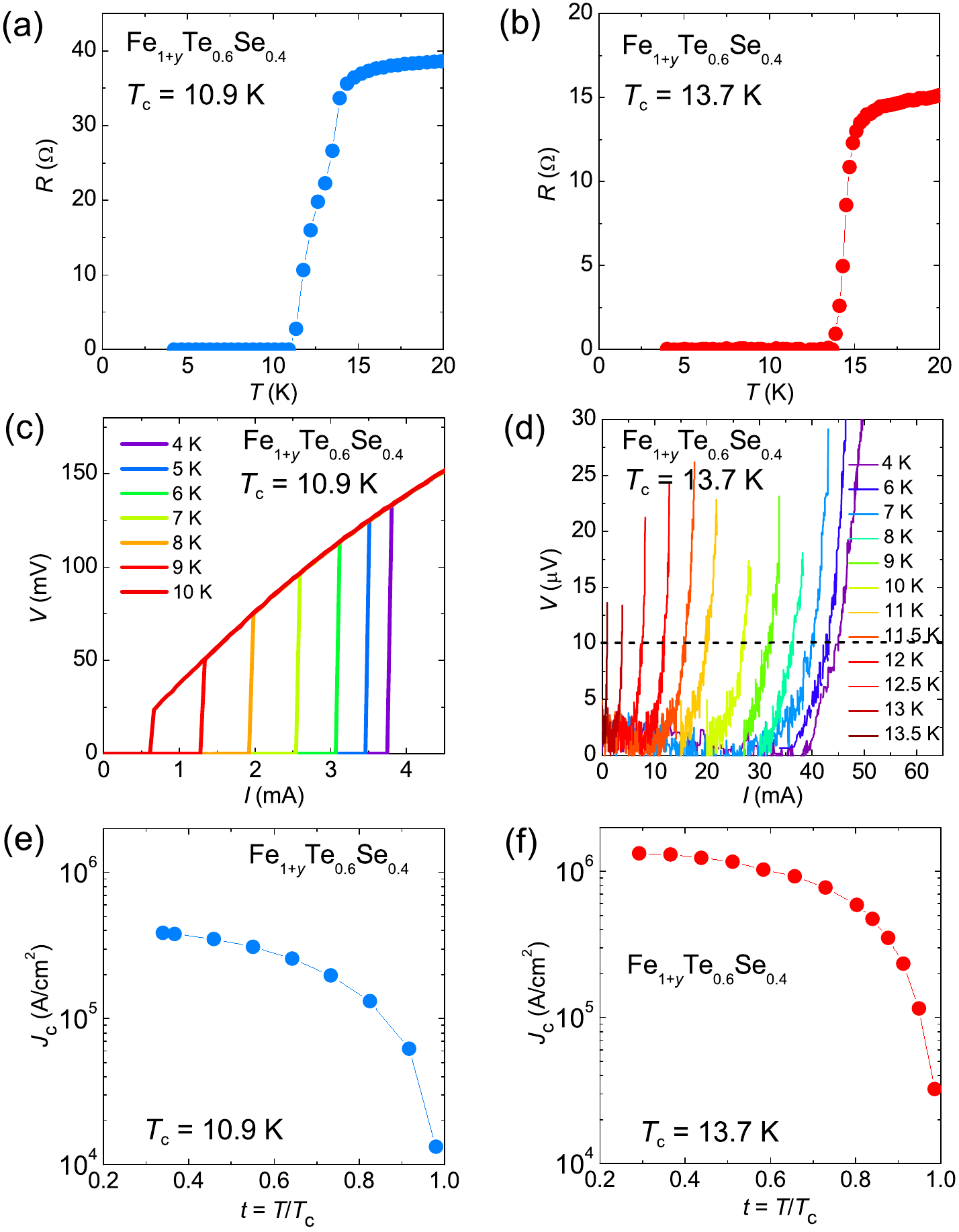}
	\caption{Temperature dependence of the resistance measured at zero field for the micro-fabricated (a) Fe$_{1+y}$Te$_{0.6}$Se$_{0.4}$ with $T_{\rm{c}}$ = 10.9 K and (b) 13.7 K, respectively. (c)-(d) The corresponding $I$-$V$ curves measured at different temperatures for the two samples. (e)-(f) Temperature dependence of the $J_{\rm{c}}$ for the two samples estimated from (c)-(d).}\label{}
\end{figure}

Figures 2(a) and 2(b) show the temperature dependence of the resistance for two typical micro-fabricated crystals with $T_{\rm{c}}$ = 10.9 K and 13.7 K, respectively. The value of $T_c$ is determined by the zero resistance. According to our previous studies, the increase of $T_{\rm{c}}$ is due to the reduction of excess Fe \cite{SunSciRep,OkadaJJAP}. The well-annealed crystal without excess Fe (e.g. the crystal with $T_{\rm{c}}$ = 13.7 K) shows a sharp SC transition width as shown in Fig. 2(b), while the crystals with some amount of left excess Fe show slightly broader transition (e.g. the crystal with $T_{\rm{c}}$ = 10.9 K shown in Fig. 2(a)). This transition suggests the slight inhomogeneity, which is derived from the residual excess Fe or the small damage due to the FIB fabrication in the narrow bridge \cite{Mollreview,Kakizaki_JJAP}. However, as described in detail below, such a broad transition does not affect the determination of the transport $J_{\rm{c}}$ in the $I$-$V$ measurements performed below $T_{\rm{c}}$. 
	
The $I$-$V$ curves for the two samples measured at different temperatures are presented in Figs. 2(c) and 2(d). The $I$-$V$ characteristics in Fig. 2(c) were measured by using the first method. The critical current, $I_{\rm{c}}$, was simply determined by the current where the voltage abruptly jumps from zero to a finite value. When the current sweeps up and down, hysteresis loops in the $I$-$V$ characteristics are observed, which are due to the Joule heating effect (see Supplemental Material \cite{supplement}). Such heating effects only occur when the applied current exceeds the critical current, which will not affect the determination of $I_{\rm{c}}$. To minimize the degradation of $I_{\rm{c}}$ by repeating the relatively large pulse current, the $I$-$V$ characteristics in Fig. 2(d) was measured by using the second method. A criterion of 10 $\mu$V (indicated as the dashed line in Fig. 2(d)) was used to define the critical current. In both methods, we emphasize that the $I_{\rm{c}}$ is determined by the SC current passing through a minimum cross-sectional area, $S$, of the bridge in the zero-resistance state below $T_{\rm{c}}$, since the SC current flows avoiding the non-SC region in the bridge. Thus, the SC transition region showing a finite resistance above $T_{\rm{c}}$ is clearly out of the scope of the $I_{\rm{c}}$ measurements performed below $T_{\rm{c}}$. Furthermore, although the effective value of $S$ can be changed by the secondary SC transiton of the non-SC region in the bridge, we confirmed that the effective $S$ for both samples was independent of temperature below $T_{\rm{c}}$, based on the following two facts. One is the smooth increase of $J_{\rm{c}}$ given by $I_{\rm{c}}/S$ shown in Figs. 2(e) and 2(f). The other is the excellent agreement between the scaled $J_{\rm{c}}$ for both crystals with $T_{\rm{c}}$= 10.9 K and 13.7 K (see Fig. 3). Thus, we conclude that the broader SC transition observed in Fig. 2(a) has no influence on the behavior of $J_{\rm{c}}$.          

\begin{figure}\center
	\includegraphics[width=8.5cm]{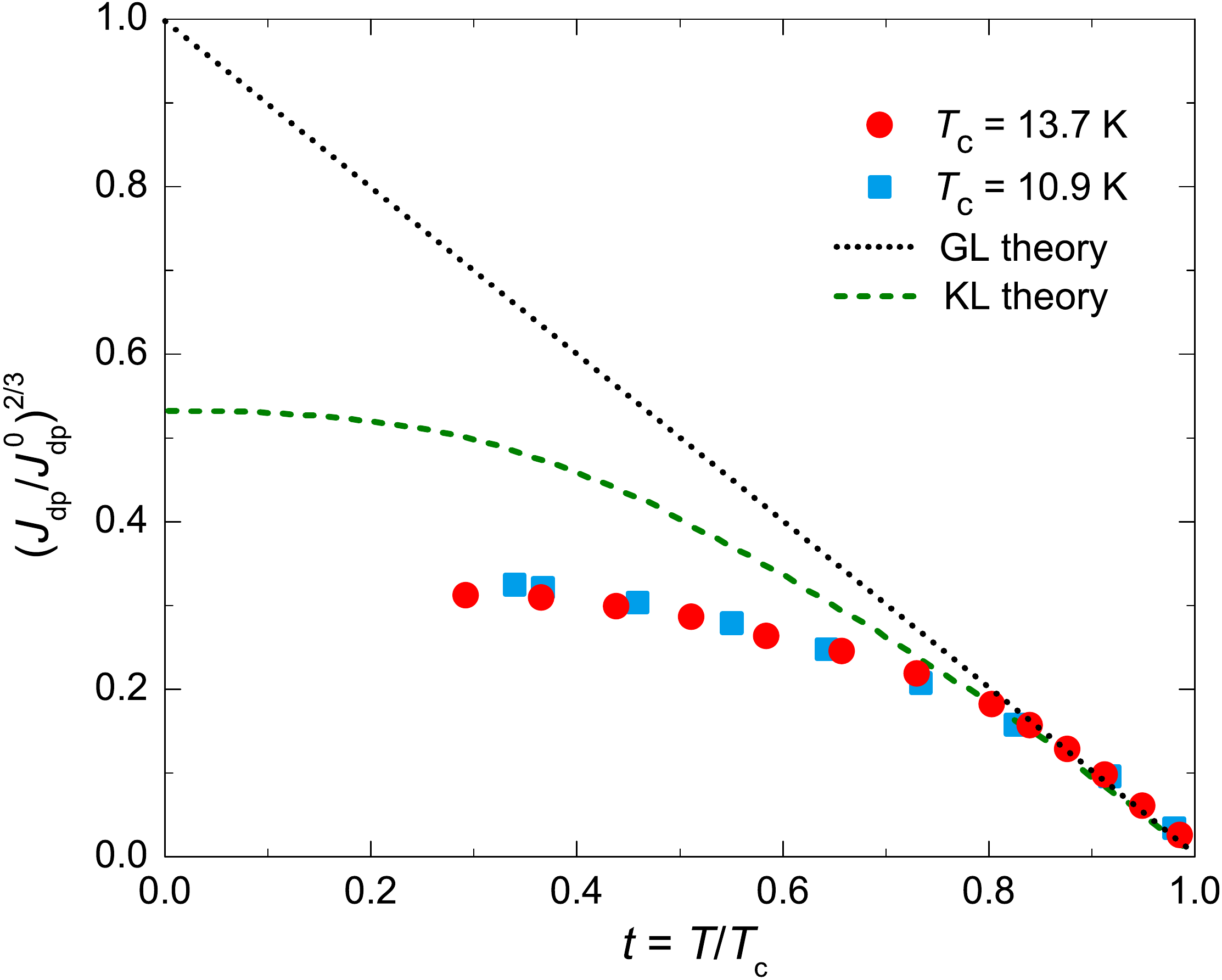}
	\caption{Reduced temperature ($t$ = $T$/$T_c$) dependence of the depairing current density, normalized to the extrapolated value $J_{\rm{dp}}^0$, along $c$-axis for the Fe$_{1+y}$Te$_{0.6}$Se$_{0.4}$ with $T_{\rm{c}}$ = 13.7 K and 10.9 K. The dotted and the dashed line represent the results from the GL-theory and KL-theory, respectively.}\label{}
\end{figure}

As shown in Figs. 2(e) and 2(f), the values of $J_{\rm{c}}$ are obviously enhanced with the increase of $T_{\rm{c}}$. In the crystal with $T_{\rm{c}}$ = 13.7 K (see Fig. 2(f)), $J_{\rm{c}}$ reaches a very large value $\sim$ 1.3 $\times$ 10$^6$ A/cm$^2$ at 4 K. This value of the transport $J_{\rm{c}}$ is about one order of magnitude larger than the depinning $J_{\rm{c}}$ ($\sim$ 1.3 $\times$ 10$^5$ A/cm$^2$ at 4 K), which has been estimated from the MHLs for both $H\parallel c$ and $H\parallel ab$ \cite{SunAPEX,Sun_SUSTevolution}. On the other hand, the depairing $J_{\rm{c}}$ along $c$-axis given in the GL-theory by $J_{\rm{dp}}^{\rm{GL}}=c\phi_0/12\sqrt3\pi^2\xi_{\rm{c}}\lambda_{\rm{c}}^2$ is estimated as $\sim$ 2$\times$10$^6$ A/cm$^2$ using the coherence length $\xi_{\rm{c}}$(0 K) $\simeq$ 3.0 nm \cite{LeiHechangPhysRevcoherencelenth}, and penetration depth $\lambda_{\rm{c}}$(0 K) $\simeq$ 1.32 $\mu$m \cite{BendelePhysRevB_lambda}. The calculated $J_{\rm{dp}}^{\rm{GL}}$(0 K) is comparable to our experimental value $\sim$ 1.3$\times$10$^6$ A/cm$^2$ at 4 K. 

In general, the critical current density at zero applied magnetic field is determined by the edge barrier for current-induced vortices, which prevents the vortex from entering the bridge. The edge barrier decreases with increasing the applied current, and will be completely suppressed when the current density approaching the critical limit, where the vortex begins to enter the bridge \cite{Tinkham}. Theoretically, the depairing current density can be obtained when the transverse dimension of the bridge is made small compared to both the coherence length and the penetration depth \cite{Tinkham}. In high-$T_{\rm{c}}$ superconductors, the condition is hard to be fulfilled since their coherence length is too small, typically several nanometers. In practice, the depairing limit can be achieved if the current is homogeneously distributed in the bridge \cite{Tinkham,NawazPhysRevLettYBCOJdp}. However, the supercurrent tends to pile up at the edges of the bridge because the magnetic flux density is the largest there as the flux lines circle the bridge \cite{SkocpolPhysRevB.14.1045,Tinkham}. This effect makes the current density nonuniform. In this case, the vortex begins to enter the bridge from edges when the averaged current density is still much smaller than the depairing limit. In other words, the obtained $J_{\rm{c}}$ is much smaller than the depairing $J_{\rm{c}}$. 

This effect has been overcome when the width of the bridge ($w$) is reduced to the Pearl length $\varLambda$ = 2$\lambda^2/h$, where $\lambda$ is the penetration depth and $h$ is the thickness of the bridge \cite{ClemPearllength}. In this case, the current density is uniformly distributed in the bridge, and the depairing limit can be achieved. Such a method has been successfully applied to achieve the depairing limit in the YBa$_2$Cu$_3$O$_{7-\delta}$ and Ba$_{0.5}$K$_{0.5}$Fe$_2$As$_2$ \cite{NawazPhysRevLettYBCOJdp,LiJunAPLdeparingJcBaK122}. In the present study, the size and the calculated Pearl length $\varLambda$ for the $c$-axis bridges are listed in Table \Romannum{2} (To directly compare with the experiments which are down to 4 K, the penetration depth at 5 K rather than 0 K is used for calculation.). The width ($w$) of the bridge is smaller than the Pearl length $\varLambda$, and comparable to the $\lambda_c$ for all the fabricated $c$-axis bridges, which guarantees the uniform current density. Thus, the large transport $J_{\rm{c}}$ observed in the $c$-axis bridge of Fe$_{1+y}$Te$_{1-x}$Se$_{x}$ single crystal is attributed to the depairing current limit. In the following paragraphs, we use $J_{\rm{dp}}$ to express the depairing current density.  

In the GL theory, $J_{\rm{dp}}$ close to $T_{\rm{c}}$ can be described as a function of the reduced temperature $t$ = $T$/$T_{\rm{c}}$ by the formula $J_{\rm{dp}}^{\rm{GL}}$($t$)=$J_{\rm{dp}}^{\rm{GL}}$(0)(1-$t$)$^{3/2}$ \cite{Tinkham}. To compare our experimental data with the theory, we fit the linear behavior of $J_{\rm{dp}}^{\rm{2/3}}$ close to $T_{\rm{c}}$, and extrapolate it to $t$ = 0 to extract $J_{\rm{dp}}^{\rm{0}}$ as used in the previous reports \cite{RusanovPhysRevB.70.024510,CirilloPhysRevB.75.174510}. The reduced temperature dependences of the normalized depairing current density ($J_{\rm{dp}}$/$J_{\rm{dp}}^{\rm{0}}$)$^{2/3}$ are shown in Fig. 3, together with the theoretical results from the GL-theory (dotted line) and KL-theory (dashed line) \cite{KLtheory}. We find that ($J_{\rm{dp}}$/$J_{\rm{dp}}^{\rm{0}}$)$^{2/3}$ for the crystal of $T_{\rm{c}}$ $\sim$ 10.9 K falls onto an identical curve for the crystal of $T_{\rm{c}}$ $\sim$ 13.9 K, confirming that $J_{\rm{c}}$ is homogeneous in the bridge part in spite of the broader resistive transition. The experimental data follows the GL-behavior, which increases linearly with decreasing $t$ down to $\sim$ 0.83 $T_{\rm{c}}$. Then, ($J_{\rm{dp}}$/$J_{\rm{dp}}^{\rm{0}}$)$^{2/3}$ gradually deviates from the GL-behavior with a reducing slope at lower $t$. Such saturation behavior was similar to those observed in low temperature superconductors \cite{SkocpolPhysRevB.14.1045,RomijnPhysRevB.26.3648,RusanovPhysRevB.70.024510,CirilloPhysRevB.75.174510}, and can be qualitatively described by the microscopic KL-theory, where the $J_{\rm{dp}}$ was numerically calculated from the Eilenberger equations by assuming that the velocity of supercurrent is proportional to a phase gradient of the SC order parameter \cite{KLtheory}.  

Quantitatively, the values of ($J_{\rm{dp}}$/$J_{\rm{dp}}^{\rm{0}}$)$^{2/3}$ are smaller than the theoretical ones at low temperatures. The fact that ($J_{\rm{dp}}$/$J_{\rm{dp}}^{\rm{0}}$)$^{2/3}$ of the two Fe$_{1+y}$Te$_{0.6}$Se$_{0.4}$ bridges with different $T_{\rm{c}}$ almost fall into an identical curve suggests the unified origin of the smaller value rather than the impurity level of the sample. The smaller $J_{\rm{dp}}$ than the theoretical one has also been observed in the micro-fabricated YBa$_2$Cu$_3$O$_{7-\delta}$ \cite{NawazPhysRevLettYBCOJdp} and some Nb thin films \cite{RusanovPhysRevB.70.024510}. The former one is attributed to the current crowding at the inner corners of the junction between the bridge and electrode, which makes the local current density at the inner corners larger than the averaged one at the center of the bridge \cite{NawazPhysRevLettYBCOJdp}. The later one is discussed to be the heating effect on the contacts, which cause extra vortex flowing \cite{RusanovPhysRevB.70.024510}. Besides, the GL-theory and KL-theory are all based on the conventional superconductors with one band structure. A recent theoretical calculation shows that the depairing $J_{\rm{c}}$ for the real material should be smaller than that expected from the KL-theory due to the broadening of the peaks in the quasi-particle density of states \cite{KuboarXiv}. Moreover, Fe$_{1+y}$Te$_{1-x}$Se$_{x}$ is a multiband system with strong inter-band scattering \cite{ChenPhysRevB.81.014526}. Future efforts of theoretical consideration on the multiband effect may be required to understand the temperature dependence of the depairing $J_{\rm{c}}$ property.        

\begin{figure}\center
	\includegraphics[width=8.5cm]{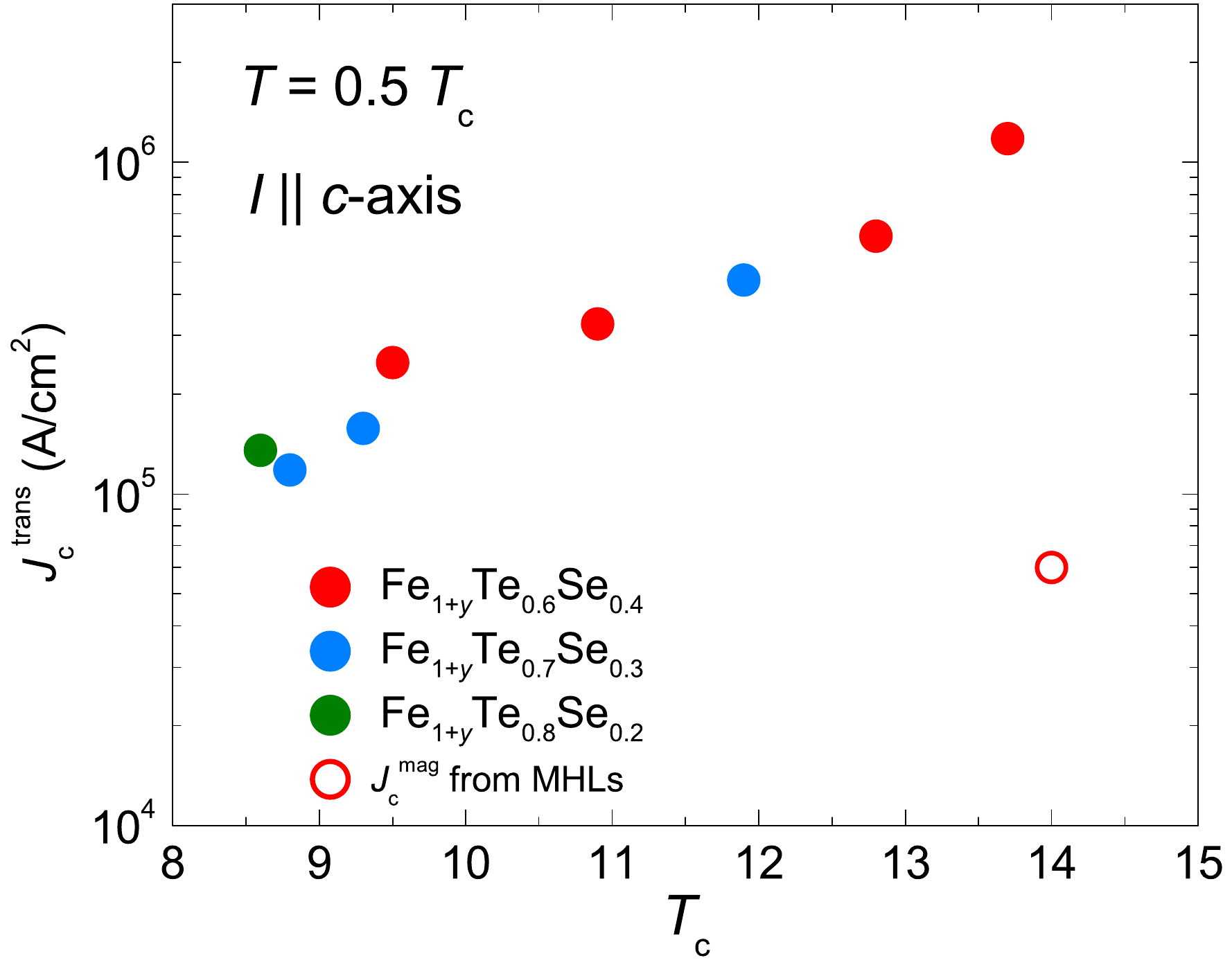}
	\caption{Evolution of $J_{\rm{c}}$ at $T$ = 0.5 $T_c$ with $T_{\rm{c}}$ for different crystals. The red, blue, and green circles represent the results obtained from the fabricated Fe$_{1+y}$Te$_{0.6}$Se$_{0.4}$, Fe$_{1+y}$Te$_{0.7}$Se$_{0.3}$, and Fe$_{1+y}$Te$_{0.8}$Se$_{0.2}$ single crystals, respectively. The open circle is the self-field $J_{\rm{c}}$ estimated from the magnetic hysteresis loops by Bean model \cite{SunAPEX,Sun_SUSTevolution}.}\label{}
\end{figure}

To directly observe the evolution of transport $J_{\rm{c}}$ with the value of $T_{\rm{c}}$, (i.e. the amount of excess Fe), we summarized the $J_{\rm{c}}$ for all the crystals at $T$ = 0.5 $T_c$, and plotted versus their values of $T_{\rm{c}}$ in Fig. 4. It is clear that $J_{\rm{c}}$ increases monotonically with $T_{\rm{c}}$ independent of the amount of Se. Note that $J_{\rm{c}}$ of the crystal with minimum $T_{\rm{c}}$ is still larger than the depinning $J_{\rm{c}}$ estimated from MHLs (See the open symbol in Fig. 4) \cite{SunAPEX,Sun_SUSTevolution}. It indicates that the depairing current limit is achieved in all the samples. On the other hand, the depairing $J_{\rm{c}}$ decreases quickly with the suppression of $T_{\rm{c}}$. Such a strong decrease of depairing $J_{\rm{c}}$ cannot be explained by the difference in bridge dimensions, since the sample with the highest $J_{\rm{c}}$ has the largest width close to the Pearl length. The decrease of depairing $J_{\rm{c}}$ may be due to the effect of excess Fe. Density functional study shows that the excess Fe is strongly magnetic, providing local moments, which will act as a pair breaker \cite{ZhangLijunPRB}. In addition, the edge barriers in the disordered region around the excess Fe is considered to be weakened, which will cause local suppression of $J_{\rm{c}}$. Our results suggest that removing the excess Fe is crucial for the increase of the current capacity limit as well as the depairing $J_{\rm{c}}$ for the Fe$_{1+y}$Te$_{1-x}$Se$_{x}$, which is instructive for other studies on thin films, tapes, and the nanoscale devices such as the single photon detectors and the nanoSQUIDs.

\section{conclusions}

In conclusion, we investigated the transport critical current density along $c$-axis in Fe$_{1+y}$Te$_{1-x}$Se$_x$ single crystals. A series of crystals containing different amounts of excess Fe with $T_{\rm{c}}$ ranging from 8.6 K to 13.7 K were fabricated into $c$-axis bridges for the transport $I$-$V$ measurements. $J_{\rm{c}}$ reaches a much larger value than the magnetic $J_{\rm{c}}$ obtained from MHLs, which is explained by reaching the depairing limit. The depairing $J_{\rm{c}}$ follows the GL-theory down to $\sim$ 0.83 $T_{\rm{c}}$, then increases with a reducing slope at lower temperatures, which can be qualitatively described by the KL-theory. This work indicates that the depairing current limit of high-$T_{\rm{c}}$ superconductors can be explored by fabricating the $c$-axis narrow bridges. Future efforts on fabricating other IBSs are expected to reveal the common behavior of the depairing $J_{\rm{c}}$ at the whole temperature range.     

\acknowledgements
The present work was partly supported by KAKENHI (JP20H05164, 19K14661, 18K03547, 16K13841, 18H05853, and 17H01141) from JSPS. FIB microfabrication performed in this work was supported by Center for Instrumental Analysis, College of Science and Engineering, Aoyama Gakuin University.


\bibliography{references}% Produces the bibliography via BibTeX.

%%%%%%%%%% Merge with supplemental materials %%%%%%%%%%
%%%%%%%%%% Merge with supplemental materials %%%%%%%%%%
\pagebreak

\newpage
%\widetext
\onecolumngrid
\begin{center}
	\textbf{\huge Supplemental information}
\end{center}
\vspace{1cm}
\twocolumngrid
%%%%%%%%%% Merge with supplemental materials %%%%%%%%%%
%%%%%%%%%% Prefix a "S" to all equations, figures, tables and reset the counter %%%%%%%%%%
\setcounter{equation}{0}
\setcounter{figure}{0}
\setcounter{table}{0}
\setcounter{page}{1}
\makeatletter
\renewcommand{\theequation}{S\arabic{equation}}
\renewcommand{\thefigure}{S\arabic{figure}}
%\renewcommand{\bibnumfmt}[1]{[S#1]}
%\renewcommand{\citenumfont}[1]{S#1}
%%%%%%%%%% Prefix a "S" to all equations, figures, tables and reset the counter %%%%%%%%%%

\section{Hysteresis loops in the I-V characteristics}

\begin{figure}\center
	\includegraphics[width=8.5cm]{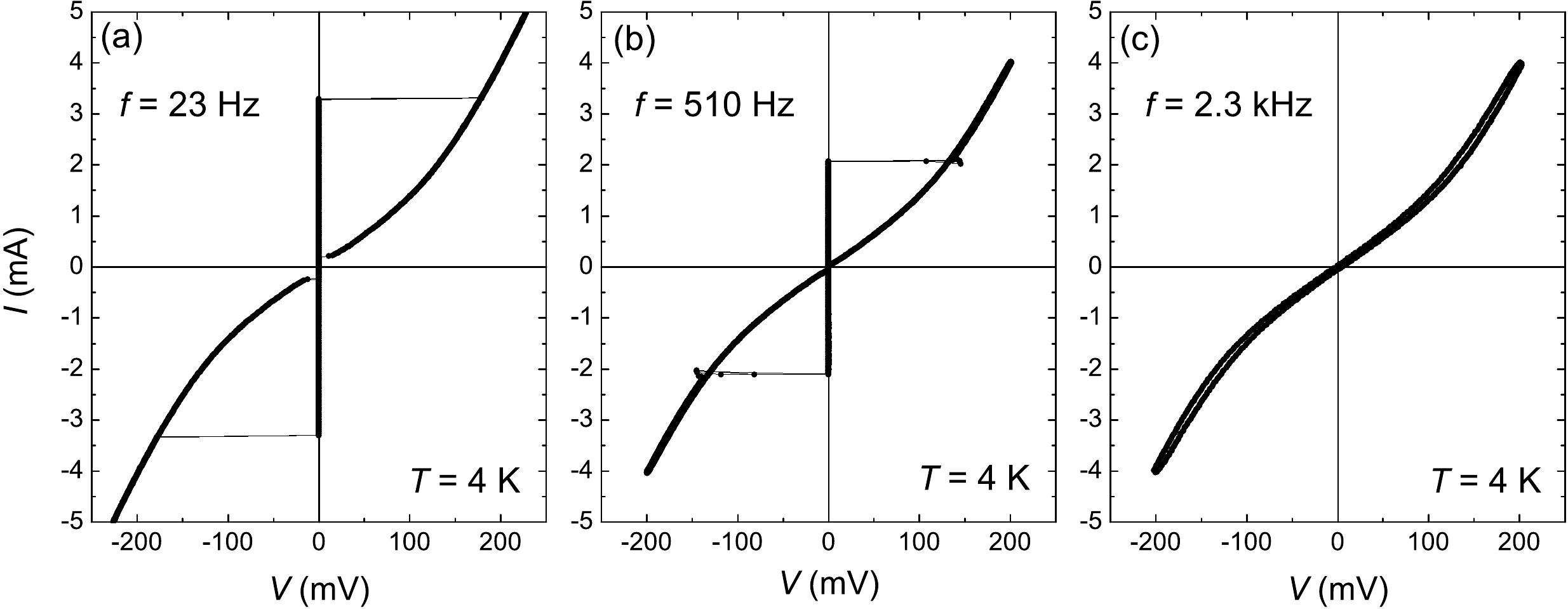}\\
	\caption{The $I$-$V$ curves of the $c$-axis bridged Fe$_{1+y}$Te$_{1-x}$Se$_x$ at 4 K measured by ramped current with frequencies of (a) 23 Hz, (b) 510 Hz, and (c) 2.3 kHz.}\label{}
\end{figure}

\begin{figure*}\center
	\includegraphics[width=18cm]{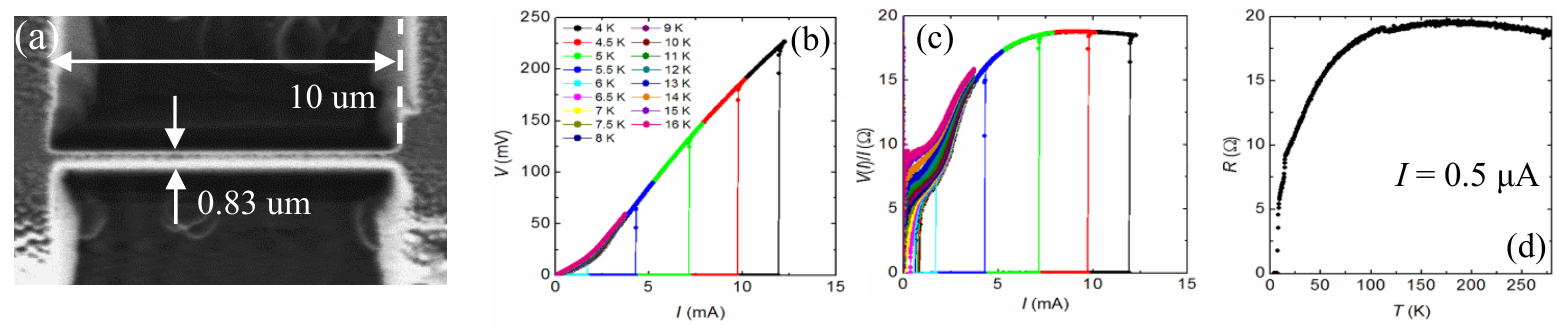}\\
	\caption{The scanning ion microscopy image of the fabricated in-plane long bridge of Fe$_{1+y}$Te$_{1-x}$Se$_x$ single crystal. (b) The $I$-$V$ curves of the in-plane long bridge measured at different temperatures. (c) The plot of $V$/$I$($\Omega$) versus $I$ for the long bridge. (d) Temperature dependence of the DC resistance for the whole sample. }\label{}
\end{figure*}

The $I$-$V$ curves of the $c$-axis bridged Fe$_{1+y}$Te$_{1-x}$Se$_x$ measured by the ramped bias current sometimes manifest hysteresis loops, which are reminiscent of the feature of underdamped Josephson junctions, as reported previously [62]. Figure S1(a) shows a typical hysteretic $I$-$V$ curve measured at 4 K with the repetition frequency of 23 Hz on the crystal with $T_{\rm{c}}$ $\sim$ 10 K. When ramping up the current, the voltage jumps from zero to a finite value at 3.2 mA, however, the voltage does not turn back to zero until the current ramping down to a much smaller value $\sim$ 0.2 mA. We found that the size of hysteresis loop is strongly dependent on the repetition frequency of the ramped bias current, as shown in Figs. S1(a) to S1(c). The diminishing behavior of the hysteresis with increasing frequencies strongly suggest that the observed hysteretic behavior is attributed to the Joule heating effect, occurring in the finite voltage state above the critical current $I_{\rm{c}}$. If the frequency is enough low, the sample will be heated up above $I_{\rm{c}}$ and be cooled down during ramping down the current, producing a large hysteresis in the $I$-$V$ curve. On the other hand, at higher frequencies, the zero voltage state disappears since there is no time period where the sample is cooled down, showing no hysteresis in the $I$-$V$ curve. 

To confirm the Joule heating effect, we micro-fabricated an in-plane bridge with length $\sim$ 10 $\mu$m, and cross-section area $\sim$ 15.6 $\mu$m$^2$. The scanning ion microscopy image from the top view is shown in Figure S2(a). Such a long bridge of superconductors must show no Josephson effect. However, the $I$-$V$ curves also show hysteresis loops, as shown in Fig. S2(b). We replot the $I$-$V$ curves into $V$/$I$ versus $I$ in Fig. S2(c), where $V$/$I$($\Omega$) represents the chordal resistance of the sample in the finite voltage state. The current dependence of $V$/$I$($\Omega$) was found to be very similar to the temperature dependence of the DC resistance (see Fig. S2(d)), confirming that the observed hysteresis in the $I$-$V$ curves are caused by the Joule heating effect. Similar hysteretic behavior in the $I$-$V$ curves has also been observed in micro-fabricated Ba$_{1-x}$K$_x$Fe$_2$As$_2$ [29]. Thus, we conclude that the superconducting state of Fe$_{1+y}$Te$_{1-x}$Se$_x$ does not belong to the intrinsic Josephson junction systems, in contrast to the (V$_2$Sr$_4$O$_6$)Fe$_2$As$_2$ [48].

\end{document}